# First Steps on Gamification of Lung Fluid Cells Annotations in the Flower Domain

Sonja Kunzmann[1], Christian Marzahl[1], Felix Denzinger[1], Christof Bertram[3], Robert Klopfleisch[4], Katharina Breininger[2], Vincent Christlein[1], Andreas Maier[1]

[1]Pattern Recognition Lab, FAU Erlangen-Nürnberg, Erlangen, Germany
[2]Department Artificial Intelligence in Biomedical Engineering, FAU Erlangen-Nürnberg, Erlangen, Germany
[3]Institute of Pathology, University of Verterinary Medicine, Vienna, Austria
[4]Institute of Veterinary Pathology, Freie Universität Berlin, Germany
Sonja.Kunzmann@fau.de

**Abstract.** Annotating data, especially in the medical domain, requires expert knowledge and a lot of effort. This limits the amount and/or usefulness of available medical data sets for experimentation. Therefore, developing strategies to increase the number of annotations while lowering the needed domain knowledge is of interest. A possible strategy is the use of gamification, i.e. transforming the annotation task into a game. We propose an approach to gamify the task of annotating lung fluid cells from pathological whole slide images (WSIs). As the domain is unknown to non-expert annotators, we transform images of cells to the domain of flower images using a CycleGAN architecture. In this more assessable domain, non-expert annotators can be (t)asked to annotate different kinds of flowers in a playful setting. In order to provide a proof of concept, this work shows that the domain transfer is possible by evaluating an image classification network trained on real cell images and tested on the cell images generated by the CycleGAN network (reconstructed cell images) as well as real cell images. The classification network reaches an average accuracy of 94.73 % on the original lung fluid cells and 95.25 % on the transformed lung fluid cells, respectively. Our study lays the foundation for future research on gamification using CycleGANs.

## 1 Introduction

On a number of tasks, computer vision algorithms have demonstrated human-level performance, especially in the areas of object detection and classification [1]. One major reason for this is the growing amount of labeled data sets which synergizes well with the advent of increasingly powerful deep learning (DL) algorithms. However, these data sets usually need to be enriched with information about their content. Most commonly performed by human annotators, this is a very time consuming, subjective and tedious task, which is also expensive and often needs prior training. Especially in the medical domain, i.e. for annotating cells in pathological whole slide images (WSIs), expert annotators usually have to get several years of practical experience. In other areas, where this constraint is not as severe, crowdsourcing is a popular approach to obtain label information. Usually crowdsourcing annotations is performed by hosting data on an interactive platform like the EXACT online annotation tool [2], where multiple annotators can assess the data simultaneously. EXACT hosts a partially annotated WSI data set [3] for the cyto-pathological diagnosis of equine asthma. However, large regions on multiple WSIs are still unlabeled. In order to overcome the issue, we propose the use of gamification approaches where game elements are applied in a non-gaming context [4]. Gamification aims to provide users with an exciting and rewarding annotation experience. An exemplary collection of successfully applied gamification in the medical domain is part of the Medical Data Donors project [5].

In order to decrease the necessary domain knowledge for untrained annotators, we propose to transform cells into flower images. The task of differentiating between various types of flowers should be possible for many human annotators. In the center of flower images the blossom is usually displayed, with a more or less homogeneous background and possibly more flowers in the background. This corresponds very well to the image perception in the cell domain where a cell is also usually located in the center of the image and other incomplete cells may be present at the edges. Figure 2 demonstrates our analogy. As a gamification strategy for the task of annotating cells, we propose the following approach (Fig. 1): in order to perform the domain transfer from cell to flower domain, we leverage the CycleGAN architecture [6]. Successful applications are e.g. style transfer, object transfiguration and season transfer as presented in Zhu at el. [6] and Bousmalis at el. [7]. As a proof of concept, we apply one CycleGAN model to each single cell type to create flower images. In the annotation game setting, the user would only see the flower image, which originates from a cell, and gets the task to annotate it in a playful way. Embodiments of the game are open for creativity but may include different tasks for different flowers (cut all sunflowers, water all daisies, etc.). In order to evaluate the performance of the CycleGAN – and therefore show that this domain transfer is possible – we evaluate cycle consistency by checking the performance of a cell type classification network using the transformed data created by transforming flowers back to the cell domain.

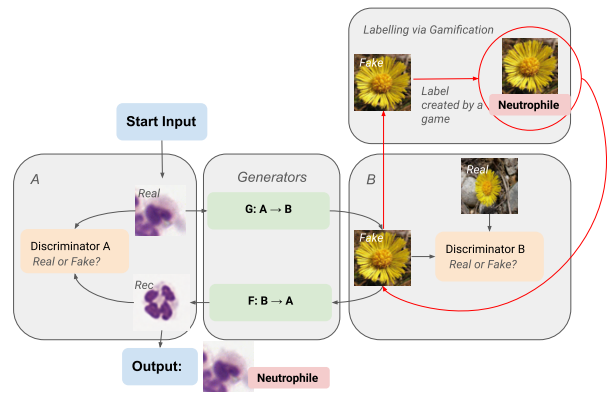

**Fig. 1.** Applying gamification for the task of annotating lung fluid cells: a CycleGAN architecture allows transformation of real cell images to reconstructed flower images and vice versa.

## 2 Materials and methods

Before going into details about the architecture and experiment design, we will present the characteristics of the two data sets used in this study, one containing lung fluid cell data and the other one flower images.

### 2.1 Lung fluid cells data set

The lung fluid cell data set consists of six cytological samples of equine bronchoalveolar lavage fluid (BALF), which were cytocentrifugated and stained using May Grunwald Giemsa. This data set was provided by Marzahl et al. [3]. Afterwards, the glass slides were digitized using a linear scanner (Aperio ScanScope CS2, Leica Biosystems, Germany) at a magnification of 400x with a resolution of 0.25 $\mu$m/px [3]. Two of the six WSIs are fully annotated and four partially annotated by a trained pathologist. The 87,738 bounding box annotated cells include neutrophils (12,556), multinuclear cells (310), mast cells (1,553), macrophages (24,498), lymphocytes (46,397), erythrocytes (339) and eosinophiles (105) [3]. Example images for the individual cell types can be seen in the top row of Figure 2. For subsequent processing, the single cells were cropped from the WSIs.

### 2.2 Flowers data set

The F17 Category Flower Data set [8] contains images of 17 different kinds of flowers with 80 images for each class. For our method, we took seven random flower types to match with a corresponding lung fluid cell type. The seven classes are coltsfoot, buttercup, daisy, windflower, daffodil, crocus, and sunflowers, as displayed in the bottom row of Figure 2. The images have large scale, pose and light variations, and there are also classes with significant variations of images within the class and close similarity to other classes [8].

### 2.3 Image transformation

In order to transform the images to the respective other domain, we decided to use a Generative Adversarial Network (GAN), which includes a generator and a discriminator and use an adversarial loss to learn the mapping such that the translated images are interchangeable with the images in the target domain. One approach is the CycleGAN architecture [6] which is an extension of the GAN architecture and involves simultaneous training of two generator and two discriminator. It captures unique characteristics of one image collection and learns how these characteristics could be transformed into another domain without paired input-output examples. Also, the CycleGAN makes further use of cycle consistency. So, that the output of the first generator propagated through the second generator should create a fake image that matches the original image. The discriminators try to classify whether the image is real or fake. Few disadvantages are information loss and incorrect conversion.

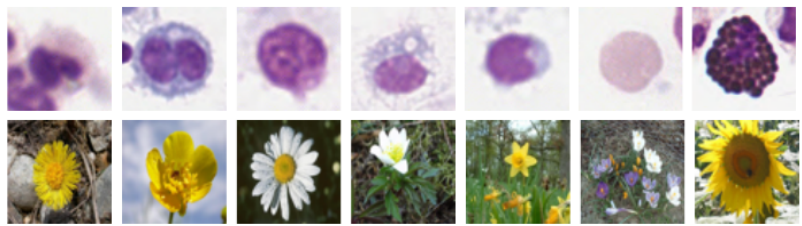

**Fig. 2.** Pairs of lung fluid cells (top) and flowers (bottom) from left to right: neutrophils and coltsfoot, multinuclear and buttercup, mast cells and daisy, macrophages and windflower, lymphocytes and daffodil, erythrocyte and crocus, eosinophile and sunflowers.

Since we first want to test whether the lung fluid cells can be converted, we trained CycleGANs with cell and flower pairs as shown in Figure 2. The architecture design and implementation for the CycleGAN is based on a publicly available Github repository (available at `https://github.com/junyanz/pytorch-CycleGAN-and-pix2pix`) which used an L1 loss function and Adam optimizer with a learning rate of 0.0002 for both generators and discriminators.

### 2.4  Experimental design

The first step is to find out if our approach is able to transform images from cells to flowers by using the CycleGAN architecture [6]. For that, we train each model for 200 epochs, with an input resolution of $64 \times 64$, and a batch size of 32. Both data sets with seven lung fluid cell types and seven flower types were divided into training (80 %), testing (10 %) and validation (10 %). In order to validate the capability of a successful domain transfer, we used the testing data of the original cell type images and reconstructed cell images to evaluate with a image classification network. The network should recognize the individual cells and should not notice any difference between real and reconstructed cells if the cycle consistency holds true. The image classification network is based on the ResNet18 architecture [9] which is pre-trained on ImageNet from the FastAi library. The model reached convergence after 10 epochs with a learning rate of $3 \times 10^{-3}$, batch size of 64, and a resolution size of $64 \times 64$. Additionally, random augmentations such as flip, rotation, random erasing and intensity shifts were applied on all training images of the data set. In order to tackle the severe class imbalance, cell types with a frequency below 2,000 were randomly oversampled to include 2,000 samples in the training set.

## 3  Results

Exemplary qualitative results for the CycleGAN approach are displayed in Figure 3. In the first row, the original domain is displayed, in the second row the reconstructed domain is shown, and in the last row the reconstructed cells are visualised. The individual cells were transformed to the corresponding domain as shown: neutrophils to coltsfoot, multinuclear to buttercup, mast cells to daisy, macrophages to windflower, lymphocytes to daffodil, erythrocyte to crocus, as well as eosinophils to sunflowers. While we see several convincing results for the converted flowers, e.g., for neutrophils (coltsfoot), the visual quality for others, e.g. daisy and crocus, may not be optimal yet (Figure 3). A possible explanation for this can be a suboptimal ratio of available cell to flower data.

Our classification network – which should differentiate between different cell types – was trained on real cells and tested on a real cell test set and a reconstructed cell test set. The corresponding confusion matrices (CMs) are depicted in Figure 4. When the network is trained and tested on the original cell images it is able to achieve an average accuracy of 94.73 %. While it reaches accuracies of up to 100.0 % for individual classes, the multinuclear cells are often misclassified as macrophages. These cell types are challenging to differentiate for the model, since the only difference really it the number of nuclei. Macrophages have one nuclei and mulinucleated cells have multiple. When testing on reconstructed data, the performance stays within the same range indicating that the domain transfer did not lead to a loss of information.

## 4   Discussion

Our method provides a proof of concept that domain conversion from lung fluid cells to flowers is possible (Fig. 3). This conversion may lower the hurdles of necessary expertise, time, and cost for crowdsourcing annotations. Additionally, a classification network did not find large differences between real or reconstructed images, which is confirmed by the average accuracy of 94.73 % on the real cell test set and an average accuracy of 95.25 % on the reconstructed cell test set. Currently, information about the labels of each cell type is necessary in order to exchange them to flowers. In future work, we would like to address this in an unsupervised or semi-supervised manner. Instead of using the cell type information as input, we could add another discriminator and loss function that takes the cell type as an additional target. Additionally, a better choice of flowers is also essential and we will analyze the gamification potential of other datasets. Specifically, we are interested in finding a suitable target data set that is optimized to allow users the differentiation of the source classes. As the current study focused on the technical feasibility, the development and evaluation of a suitable annotation game is an important next step towards translation.

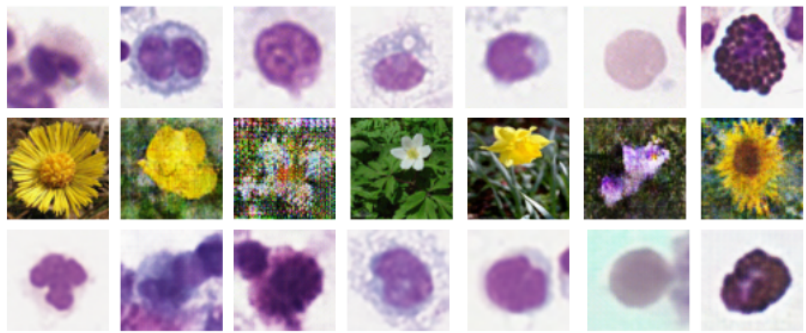

**Fig. 3.** Results of the domain transformation from real lung fluid cells (first row) to fake flowers (second row) and reconstructed lung fluid cells (third row).

**Fig. 4.** Confusion matrices for the image classification network trained on the original cell type images and tested on the real cell images (left) and tested on the reconstructed cell images (right).

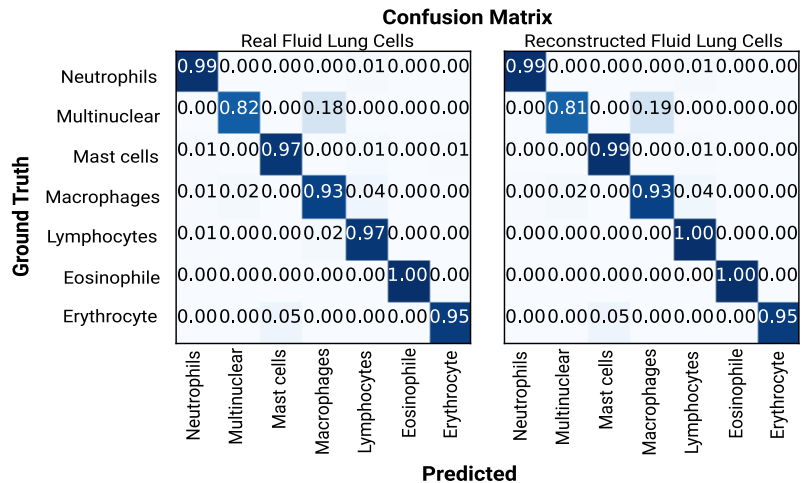